# MR Elastography with Optimization-Based Phase Unwrapping and Traveling Wave Expansion-based Neural Network (TWENN)

Shengyuan Ma, Runke Wang, Suhao Qiu, Ruokun Li, Qi Yue, Qingfang Sun, Liang Chen, Fuhua Yan, Guang-Zhong Yang, Yuan Feng

*Abstract*— **Magnetic Resonance Elastography (MRE) can characterize biomechanical properties of soft tissue for disease diagnosis and treatment planning. However, complicated wavefields acquired from MRE coupled with noise pose challenges for accurate displacement extraction and modulus estimation. Using optimization-based displacement extraction and Traveling Wave Expansion-based Neural Network (TWENN) modulus estimation, we propose a new pipeline for processing MRE images. An objective function with Dual Data Consistency (Dual-DC) has been used to ensure accurate phase unwrapping and displacement extraction. For the estimation of complex wavenumbers, a complex-valued neural network with displacement covariance as an input has been developed. A model of traveling wave expansion is used to generate training datasets for the network with varying levels of noise. The complex shear modulus map is obtained through fusion of multifrequency and multidirectional data. Validation using brain and liver simulation images demonstrates the practical value of the proposed pipeline, which can estimate the biomechanical properties with minimal root-mean-square errors when compared to state-of-the-art methods. Applications of the proposed method for processing MRE images of phantom, brain, and liver reveal clear anatomical features, robustness to noise, and good generalizability of the pipeline.**

*Index Terms*—**Magnetic resonance elastography, Modulus estimation, Neural network, Traveling waves, Phase unwrapping**

## I. INTRODUCTION

Magnetic Resonance Elastography (MRE) can measure viscoelastic mechanical parameters of soft tissues noninvasively [1], [2]. In recent years, studies have investigated the diagnostic potential of MRE for liver cirrhosis [3], brain tumors [4], liver tumors [5], and Parkinson's diseases [6]. Post-processing of MRE images includes two major steps:

1) extraction of wavefield from the original wrapped phase; and 2) estimation of biomechanical properties based on the measured wavefield [7]. However, accurate displacement extraction from wrapped phase images with noise remains difficult [8]. In vivo wavefields of soft tissues with complicated anatomical structures usually contain deflections and reflections with acquisition noise. These bring challenges to the accurate estimation of the biomechanical properties of soft tissues [9].

As the first step of processing MRE images, the quality of the extracted wavefield determines the overall performance of the extracted biomechanical properties. In phase images, wave information is encoded with a motion encoding gradient, such that the magnitude of the motion can introduce a wrapping effect. Thus, phase unwrapping is necessary. Conventional approaches involve unwrapping a single image frame before performing Fast Fourier transform (FFT) in the time domain to extract the principal component [10]. In addition, spatiotemporal information can also be used for phase unwrapping [11]. Unwrapping algorithms such as Sorting by Reliability (SG) algorithm [12] and Dilate-Erode-Propagate (DE) [8] work in a search-like or discrete greedy update manner, which can fail in complicated wrapped scenarios [8]. Laplacian-based Estimation (LBE) simultaneously performs phase unwrapping and FFT in the frequency domain [13]. LBE is noise-resilient, but may introduce additional background offset noise. This can have a negative impact on the estimated biomechanical properties. The phase gradient method (PG) does not need direct phase unwrapping, but is noise-sensitive and requires specially designed modulus estimation algorithms [14]. Therefore, a noise-resistant, accurate method for wavefield extraction that can perform both phase unwrapping and extraction of the principal component is required.

MRE is a shear-wave-based elastography that requires the estimation of complex wavenumbers from acquired wave images. Numerous methods have been proposed to date [9] and

This work was supported by grant 32271359 from the National Natural Science Foundation of China, grant 22ZR1429600 from the Natural Science Foundation of Shanghai, grant 2022YFB4702700, 2022YFB4702704 from the National Key R&D Program of China, grant 20DZ2220400 from the Science and Technology Commission of Shanghai Municipality, and grant 2021SHZDZX from the Shanghai Municipal Science and Technology Major Project. (Corresponding author: Yuan Feng.)

S. Ma, R. Wang, S. Qiu, G. Z. Yang and Y. Feng are with the School of Biomedical Engineering, Institute of Medical Robotics, and the National Engineering Research Center of Advanced Magnetic Resonance Technologies for Diagnosis and Therapy (NERC-AMRT), Shanghai Jiao Tong University, Shanghai 200030, China (e-mail: fengyuan@sjtu.edu.cn).

R. Li, F. Yan, and Y. Feng are with the Department of Radiology, Ruijin Hospital, School of Medicine, Shanghai Jiao Tong University, Shanghai, China.

Q. Sun is with the Department of Neurosurgery, Ruijin Hospital, School of Medicine, Shanghai Jiao Tong University, Shanghai, China.

Q. Yue and Liang Chen are with the Department of Neurosurgery, Huashan Hospital, Fudan University, Shanghai, China.

typical algorithms include Direct Inversion (DI) using the Helmholtz equation [15] and local frequency estimation (LFE) [16]. Due to the use of Laplace operators, DI is susceptible to noise and prone to edge artifacts. Iterative algorithms with prior information such as Multifrequency Elasticity Reconstruction using Structured Sparsity and ADMM (MERSA) [17] and MRE Inversion by Compressive Recovery (MICRo) [18] were proposed to use DI as an estimation kernel. But these methods had similar problems due to the differentiation operation. LFE is more noise-resistant, but its ability to recover anatomical features is limited. Enhanced Complex Local Frequency (EC-LFE) was developed to provide viscosity information, but structural resilience remained a problem [19].

To overcome the limitations of the DI-based method, Multifrequency Dual Elasto-Visco inversion (MDEV) was proposed to improve noise robustness using multi-frequency data averaging [20]. However, differentiation operation could still introduce structural edge artifacts, making it sensitive to noise. To improve the capability of distinguishing anatomical features, k-MDEV applied directional filter banks to suppress noise and new estimation kernels with a single-wave assumption [21]. However, the use of the Laplace operator may over-enhance the object boundary and the assumption of single-wave may be invalid when there are strong reflections. Elastography Software Pipeline (ESP) [22] used a series of filters for denoising and Gabor wavelets for inversion. ESP could recover refined anatomical details but required many parameter-tuning steps.

Non-Linear Inversion (NLI) algorithms based on Finite Element (FE) can provide good inversion [9], [23], [24]. However, the computation cost was high and the boundary condition settings could greatly influence the results [25]. Notably, by implementing a parallelized, subzone-based domain decomposition approach, the NLI algorithm proposed by the Dartmouth group avoids undue bias from the applied boundary conditions and high computational cost [26]. The method has been used to establish benchmark values for viscoelastic properties of the human brain [27]. Recently, data-driven algorithms using Deep Learning (DL) were proposed using the training set from FE simulation [28], [29]. These methods also relied on specific FE models for specific application scenarios, hindering their generalization performance. Traveling Wave Expansion (TWE) was first introduced to solve the inversion problem by using large fitting windows [30]. However, the inverse of the ill-conditioned matrix limited the spatial resolution that can be achieved. Therefore, a noise-resistant method with low-computational-cost, clearly delineated anatomical features and strong generalizability is required.

### A. Paper Contribution

In this study, we propose an optimization-based phase unwrapping method with Dual Data Consistency (Dual-DC) to simultaneously perform phase unwrapping and FFT. For the inversion of complex shear modulus, a TWE-based Neural Network (TWENN) algorithm is proposed. Detailed Simulation, phantom, and human studies were performed to demonstrate the pipeline's accuracy and reliability.

### B. Paper Organization

The remaining sections are organized as follows: In Section II, the proposed pipeline including Dual-DC and TWENN is presented. In Section III, the test datasets, evaluation metrics, implementation details of the proposed pipeline and comparative algorithms are introduced. Section IV presents detailed results followed by Section V that provides a thorough analysis of the results.

## II. METHODS

The proposed pipeline for processing MRE images includes an optimization-based phase unwrapping method with Dual-DC, and a Traveling Wave Expansion-based Neural Network (TWENN) modulus estimation to estimate complex shear modulus (Fig. 1).

### A. Optimization-based Phase Unwrapping with Dual Data Consistency

The displacement extraction of phase images acquired from MRE generally consists of two steps: phase unwrapping and computation of the principal components using FFT [7]. Here, an optimization-based method was proposed to perform these two tasks simultaneously.

#### 1) Problem Formulation

If $U^* = U' + i \cdot U''$ is the complex displacement field rotating at a frequency $\omega$, representing the oscillating motion. In the process of image acquisition, at a specific time point $t_j (j = 1, \dots, J)$ within the motion cycle where the phase offset is $\varphi_j = \omega t_j$, the image phase recorded by motion encoding gradient is $\boldsymbol{\phi}_j$:

$$\boldsymbol{\phi}_j = \text{Re}(\boldsymbol{U}^* \cdot e^{i\varphi_j}) = \boldsymbol{U}' \cdot \cos(\varphi_j) - \boldsymbol{U}'' \cdot \sin(\varphi_j), \quad (1)$$

Therefore, the image $\boldsymbol{I}_j$ acquired at $t_j$ is

$$\boldsymbol{I}_j = |\boldsymbol{A}| \cdot e^{i\boldsymbol{\Phi}} \cdot e^{i\boldsymbol{\phi}_j}, \quad (2)$$

where $|\boldsymbol{A}|$ is the magnitude of image and $\boldsymbol{\Phi}$ is the background phase of $\boldsymbol{I}_j$. The purpose of displacement extraction is to obtain $\boldsymbol{U}^*$ from $\boldsymbol{I}_j (j = 1 \dots J)$, which includes two tasks: computation of principal component and phase unwrapping. Two objective functions are proposed to solve them, respectively.

#### 2) Objective Function for Principal Components

Cross differences among the acquired images are used to remove the background phase. For images acquired at two temporal points $t_p$ and $t_q$, $(p, q) \in \{(x, y) | x < y, 1 \le x, y \le J\}$:

$$\frac{\boldsymbol{I}_p}{\boldsymbol{I}_q} = e^{i(\boldsymbol{\phi}_p - \boldsymbol{\phi}_q)}$$
$$= e^{i\{\boldsymbol{U}' \cdot [\cos(\varphi_p) - \cos(\varphi_q)] + \boldsymbol{U}'' \cdot [-\sin(\varphi_p) + \sin(\varphi_q)]\}} \quad (3)$$

Define the constant coefficients $E_{pq}$ and $F_{pq}$,

$$E_{pq} = \cos(\varphi_p) - \cos(\varphi_q)$$
$$F_{pq} = -\sin(\varphi_p) + \sin(\varphi_q) \quad (4)$$

The relationship between unknown $\boldsymbol{U}^*$ and known $\boldsymbol{I}_j$ is established so that the principal components are obtained. Thus, the first data consistency (DC) term of phase can be written as:

$$\mathcal{O}_{DC1} = \sum_{p,q} \left\| \frac{\boldsymbol{I}_p}{\boldsymbol{I}_q} - e^{i(\boldsymbol{U}' \cdot E_{pq} + \boldsymbol{U}'' \cdot F_{pq})} \right\|_2 \quad (5)$$

### 3) Objective Function for Unwrapping

With Eq. (5) only, $\boldsymbol{U}^*$ may still be wrapped. The absolute value of the wrapped phase gradient is much higher than that after unwrapping. Using the chain rule, the gradient of phase after unwrapping (true phase gradient) can be obtained directly from the wrapped phase [7]. Thus, data consistency is achieved between the true phase gradient and the gradient of $\boldsymbol{U}^*$.

If the background phase is ignored, $\frac{I_j}{|I_j|} \approx e^{i\phi_j}$, by using the chain rule, the phase gradient is calculated from the wrapped phase,

$$\frac{\partial e^{i\phi_j}}{\partial x} = i \cdot e^{i\phi_j} \cdot \frac{\partial\left(\boldsymbol{U}' \cdot \cos(\varphi_j) - \boldsymbol{U}'' \cdot \sin(\varphi_j)\right)}{\partial x} \tag{6}$$

$$\cos(\varphi_j)\frac{\partial \boldsymbol{U}'}{\partial x} - \sin(\varphi_j)\frac{\partial \boldsymbol{U}''}{\partial x} = -\frac{i}{e^{i\phi_j}}\frac{\partial e^{i\phi_j}}{\partial x}.$$

In matrix form, we have

$$\begin{bmatrix} \cos(\varphi_1) & -\sin(\varphi_1) \\ \vdots & \vdots \\ \cos(\varphi_j) & -\sin(\varphi_j) \end{bmatrix} \begin{bmatrix} \frac{\partial \boldsymbol{U}'}{\partial x} \\ \frac{\partial \boldsymbol{U}''}{\partial x} \end{bmatrix} = \begin{bmatrix} -\frac{i}{e^{i\phi_1}}\frac{\partial e^{i\phi_1}}{\partial x} \\ \vdots \\ -\frac{i}{e^{i\phi_j}}\frac{\partial e^{i\phi_j}}{\partial x} \end{bmatrix}. \tag{7}$$

Let $\boldsymbol{H} = \begin{bmatrix} \cos(\varphi_1) & -\sin(\varphi_1) \\ \vdots & \vdots \\ \cos(\varphi_j) & -\sin(\varphi_j) \end{bmatrix}$, $\begin{bmatrix} \frac{\partial \boldsymbol{U}'}{\partial x} \\ \frac{\partial \boldsymbol{U}''}{\partial x} \end{bmatrix} = \begin{bmatrix} \widetilde{\boldsymbol{U}'}_x \\ \widetilde{\boldsymbol{U}''}_x \end{bmatrix}$, then

$$\begin{bmatrix} \widetilde{\boldsymbol{U}'}_x \\ \widetilde{\boldsymbol{U}''}_x \end{bmatrix} = (\boldsymbol{H}'\boldsymbol{H})^{-1}\boldsymbol{H}' \begin{bmatrix} -\frac{i}{e^{i\phi_1}}\frac{\partial e^{i\phi_1}}{\partial x} \\ \vdots \\ -\frac{i}{e^{i\phi_j}}\frac{\partial e^{i\phi_j}}{\partial x} \end{bmatrix}. \tag{8}$$

Thus, the second data consistency term (DC) of phase gradient can be defined as

$$\mathcal{O}_{DC2} = \left\|\frac{\partial \boldsymbol{U}'}{\partial x} - \widetilde{\boldsymbol{U}'}_x\right\|_2 + \left\|\frac{\partial \boldsymbol{U}'}{\partial y} - \widetilde{\boldsymbol{U}'}_y\right\|_2 + \left\|\frac{\partial \boldsymbol{U}''}{\partial x} - \widetilde{\boldsymbol{U}''}_x\right\|_2 + \left\|\frac{\partial \boldsymbol{U}''}{\partial y} - \widetilde{\boldsymbol{U}''}_y\right\|_2. \tag{9}$$

### 4) Optimization Using Dual-DC

The displacement extraction problem in MRE including FFT and unwrapping can be treated as an optimization problem (10) by combining (5) and (9) as an objective function. Here, the adaptive Momentum (ADAM) algorithm was used to solve this optimization problem.

$$\min_{U} \left\{ \sum_{p,q} \left\|\frac{I_p}{I_q} - e^{i(\boldsymbol{U}' \cdot E_{pq} + \boldsymbol{U}'' \cdot F_{pq})}\right\|_2 + \lambda \left( \begin{array}{c} \left\|\frac{\partial \boldsymbol{U}'}{\partial x} - \widetilde{\boldsymbol{U}'}_x\right\|_2 + \left\|\frac{\partial \boldsymbol{U}'}{\partial y} - \widetilde{\boldsymbol{U}'}_y\right\|_2 \\ + \left\|\frac{\partial \boldsymbol{U}''}{\partial x} - \widetilde{\boldsymbol{U}''}_x\right\|_2 + \left\|\frac{\partial \boldsymbol{U}''}{\partial y} - \widetilde{\boldsymbol{U}''}_y\right\|_2 \end{array} \right) \right\}, \tag{10}$$

where $\lambda$ is a weighting parameter.

---

**Algorithm 1** Dual-DC phase unwrapping

**Input**: Measured complex MR images $\boldsymbol{I}_j$, Phase offset $\varphi_j$ for each image.

**Output**: Main displacement components $\boldsymbol{U}^* = \boldsymbol{U}' + i \cdot \boldsymbol{U}''$

1. Initialize the $\boldsymbol{U}^*$ to be updated and set weighting parameter $\lambda$ and maximum number of iterations $maxIter$. Set number of iterations $L = 0$.
2. Calculate cross phase differences $\frac{I_p}{I_q}$. (3)
3. Calculate coefficients $E_{pq}$ and $F_{pq}$ using $\varphi_j$. (4)
4. Calculate phase gradient $\widetilde{\boldsymbol{U}'}_x, \widetilde{\boldsymbol{U}'}_y, \widetilde{\boldsymbol{U}''}_x, \widetilde{\boldsymbol{U}''}_y$ from MR images $\boldsymbol{I}_j$. (8)
5. **repeat**
6.     Update $\mathcal{O}_{DC1}$. (5)
7.     Update the gradient of $\boldsymbol{U}^*, \frac{\partial \boldsymbol{U}'}{\partial x}, \frac{\partial \boldsymbol{U}'}{\partial y}, \frac{\partial \boldsymbol{U}''}{\partial x}, \frac{\partial \boldsymbol{U}''}{\partial y}$.
8.     Update $\mathcal{O}_{DC2}$ (9)
9.     Update $\boldsymbol{U}^*$ using ADAM algorithm. (10)
10.     $L = L + 1$
11. **until** $L \geq maxIter$
12. **return** $\boldsymbol{U}^* = \boldsymbol{U}' + i \cdot \boldsymbol{U}''$

---

### B. TWENN: Traveling Wave Expansion-based Neural Network Modulus Estimation

The TWE model was used to generate noisy training data. It was also used to construct and train a complex value covariance neural network to mine the mapping from wavefield to wavenumber, and finally use multi-frequency multi-directional fusion to obtain the complex shear modulus.

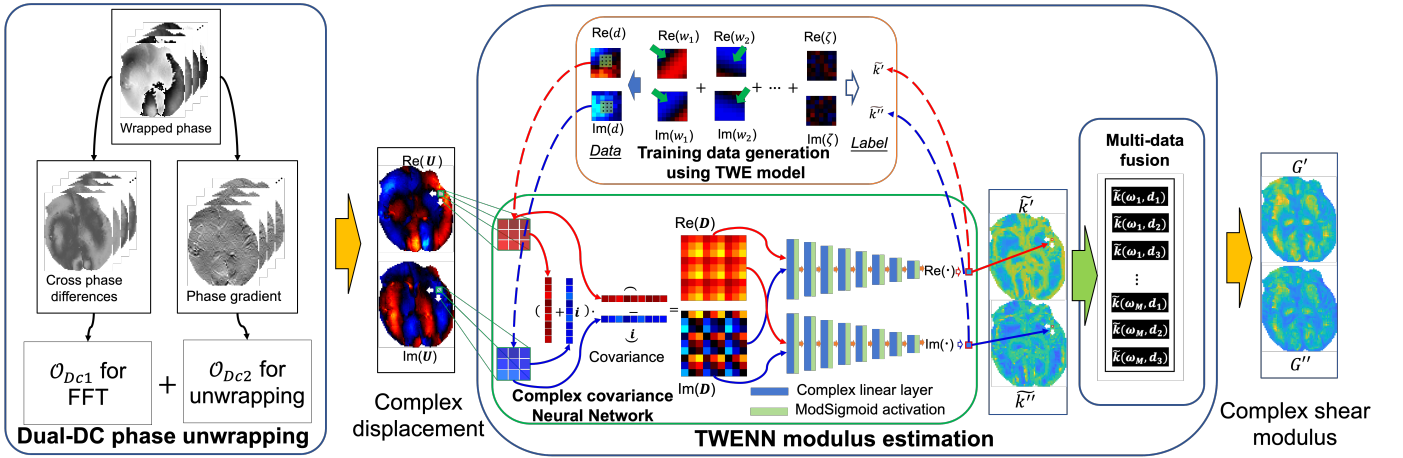

Fig. 1. A flow chart of the pipeline of MRE image processing for displacement extraction, training data generation, structure of network and multi-data fusion. Complex displacement field was extracted from phase images using an optimization of an objective function with dual-DC. Training data was generated by TWE model for network training. The normalized complex wavenumber was estimated by TWENN. Complex shear modulus was obtained by combining multi-frequency and multi-direction data.

### 1) Traveling Wave Expansion

For homogeneous, incompressible, isotropic, and viscoelastic material, at spatial location $r$, the complex value displacement is $u(r)$. It is expressed as the superposition of $M$ traveling waves [30]:

$$u(r) = \sum_{m=1}^{M} w_m$$
$$= \sum_{m=1}^{M} a_m \cdot e^{i \cdot (k' + i \cdot k'') \cdot \widehat{n_m} \cdot r} \quad (11)$$

where $w_m$ is the $m$th traveling waves, $a_m$ is the complex value amplitude of the displacement, $m$ is the number of traveling waves, $\widehat{n_m}$ is the unit vector of the propagation direction of the $m$th traveling wave, $k^* = k' + i \cdot k''$ is the local complex wavenumber, $k'$ is the real wavenumber related to material elasticity, $k''$ is imaginary wavenumber related to material viscosity. The complex wavenumber can be normalized by vibration frequency $\omega$:

$$\widetilde{k^*} = \frac{k^*}{\omega} = \frac{k'}{\omega} + i \cdot \frac{k''}{\omega}$$
$$= \widetilde{k'} + i \cdot \widetilde{k''}. \quad (12)$$

Thus, the aim of the inversion is to find the operator $\mathcal{F}_\theta: u \longrightarrow \widetilde{k^*}$.

### 2) Training Data Generation

To construct a neural network capable of realizing the mapping operator $\mathcal{F}_\theta$, training data set was prepared using TWE model. The training set was created by varying different numbers of traveling waves $M$, propagation direction $\widehat{n_m}$, complex amplitude $a_m$ and complex wavenumber $k$. In this study, an isotropic patch of 3×3 (2D) or 3×3×3 (3D) was used for the inversion of $k$ at each location. The relatively small patch used is to increase the spatial resolution. To obtain $\mathcal{F}_\theta$ with better noise robustness, noise $\zeta(r)$ was added to the training set. Suppose $\mathbb{C}(r)$ is the normalized, complex-valued Gaussian noise, $b$ is the intensity of the noise, $\zeta(r) = b \cdot \mathbb{C}(r)$. The detail of dataset generation rules is shown in Appendix. The training data $d(r)$ at location $r$ can be written as:

$$d(r) = u(r) + \zeta(r)$$
$$= \sum_{m=1}^{M} a_m \cdot e^{i \cdot (k + i \cdot k'') \widehat{n_m} \cdot r} + b \cdot \mathbb{C}(r) \quad (13)$$

For 2D cases, the training data $\boldsymbol{d}$ can be written as a 2D tensor as follows:

$$\boldsymbol{d} = \begin{bmatrix} d(r_{-1,-1}) & d(r_{0,-1}) & d(r_{1,-1}) \\ d(r_{-1,0}) & d(r_{0,0}) & d(r_{1,0}) \\ d(r_{-1,1}) & d(r_{0,1}) & d(r_{1,1}) \end{bmatrix} \quad (14)$$

Where $r_{x,y}$ is a vector pointing from the center of the patch to the target pixel position $(x, y)$. Likewise, for 3D cases, $\boldsymbol{d}$ will be a three-dimensional tensor.

### 3) Covariance Complex Neural Network

Studies have shown complex value neural networks are more suitable for solving complex-valued problems [31]. Therefore, a complex-valued neural network was constructed to process these complex-valued wavefields. In a cascade structure, TWENNs had 9-layer fully connected complex-value neural networks, with the number of neurons in each layer being

60,50,40,30,20,15,10,6,1, respectively. No extra layers were used.

If the weight matrix of the linear layer is W = A + $i$B, and the complex input is h = $\alpha$ + $i\beta$, the output is Wh = A$\alpha$ − B$\beta$ + $i$(A$\beta$ + B$\alpha$). To preserve the phase angle, and to keep the magnitude within a certain range, a new complex activation function named *modSigmoid* is designed [32]:

$$\text{modSigmoid}(z) = \text{Sigmoid}(|z| + a)e^{i\theta_z}, \quad (15)$$

where z is the complex input data, a is a real bias to be learned, $\theta_z$ is the angle of z and $\text{Sigmoid}(x) = \frac{1}{1+e^{-x}}$.

Because the covariance can expose the information within the wave field better [33], a covariance term $\boldsymbol{D} = \text{vec}(\boldsymbol{d})\text{vec}(\boldsymbol{d})^H$ is used as the model input, where $\text{vec}(\cdot)$ is an operator transforms a tensor to a column vector. $\widetilde{k'}$ and $\widetilde{k''}$ are estimated separately by two parallel multi-layer full connection networks (Fig. 1). Two separate networks are used to estimate $\widetilde{k'}$ and $\widetilde{k''}$. For each network, mean square error is used as a loss function for optimization. In this study, the network was trained using the ADAM algorithm. $\mathcal{F}_\theta(\omega)$ was trained and obtained at different frequencies $\omega$.

### 4) Mutli-frequency and Mutli-direction Fusion

For the wavefield $\boldsymbol{U}^*_{\omega_m, d_n}$ obtained from MRE at different frequencies $\omega_m (m = 1, ..., M)$ and in different directions $d_n (n = 1, ..., N)$, the corresponding normalized complex wavenumber $\widetilde{k^*}(\omega_m, d_n)$ is obtained using the trained network $\mathcal{F}_\theta(\omega_m)$.

For complex shear modulus $G^* = G' + iG''$, using corresponding principle [17],

$$G^* = \rho \frac{\omega^2}{(k' - i \cdot k'')^2} = \frac{\rho}{(\widetilde{k'} - i \cdot \widetilde{k''})^2}. \quad (16)$$

Thus, the storage modulus $G'$ and loss modulus $G''$ can be estimated from the average of $\widetilde{k^*}(\omega_m, d_n)$:

$$G' + iG'' = \frac{\rho}{\left(\frac{\sum_{m,n} \widetilde{k'}(\omega_m, d_n) - i \sum_{m,n} \widetilde{k''}(\omega_m, d_n)}{MN}\right)^2}. \quad (17)$$

This method does not do any pre-filtering to filter the noise, and the final modulus is filtered by a 3×3 median filter.

---

**Algorithm 2** Modulus estimation with TWENN

**Input**: Wavefields of multi-frequency and multi-direction $\boldsymbol{U}^*_{\omega_m, d_n}$, information of spatial resolution and vibration frequencies $\omega_m (m = 1, ..., M)$

**Output**: Complex shear modulus $G^* = G' + iG''$

1. Set the patch size, and intensity of the noise $b$.
2. Generate training data based on spatial resolution and vibration frequencies using the traveling wave expansion model. (13)
3. Construct covariance complex neural network $\mathcal{F}_\theta$. (14,15)
4. Training $\mathcal{F}_\theta(\omega)$ using ADAM.
5. Calculate $\widetilde{k^*}(\omega_n, d_m)$ using $\mathcal{F}_\theta$.
6. Estimate $G^* = G' + iG''$ using multi-frequency multi-directional data fusion. (17)
7. **return** $G^* = G' + iG''$

---

## III. Validation and verification

Wrapped phase images from simulation liver dataset and in vivo brain MRE experiment were used for the validation of the proposed phase unwrapping method with Dual-DC. Modulus estimation with TWENN was validated using simulated phantom, brain, and liver wavefields. The proposed pipeline combining Dual-DC and TWENN was validated using phantom, brain, or liver MRE images. A summary of the data sets and methods used is shown in TABLE I. Results are also compared with those from state-of-the-art methods.

### A. Validation of Dual-DC phase unwrapping

#### 1) Simulated Liver Dataset

The artificially wrapped phase was generated using 2D wavefield data for publicly available simulated liver images (https://bioqic-apps.charite.de/). The displacements were normalized with a maximum value of $4\pi$. Phases at 4 different temporal points within one cycle were generated. Complex Gaussian noise with different intensities was added to the complex image to evaluate the robustness of the algorithm.

#### 2) In vivo 3D Brain Dataset

The in vivo brain MRE images were acquired using a 3T scanner (uMR790, United Imaging Healthcare, Shanghai, China) with TR/TE=4000/65ms, MEG=40mT/m, resolution=3 mm × 3 mm × 3 mm, and 8 phase offsets [34]. The study protocol was reviewed and approved by the Institution Review Board of Shanghai Jiao Tong University.

#### 3) Algorithm Comparison

The performance of phase unwrapping method is compared with that from SR algorithm [12] and LBE [13], [35] that are commonly used in postprocessing MRE images [8]. Simulation results are compared with the Ground Truth (GT) while mean errors are calculated for validation. Interlayer continuity of the unwrapped phase is compared for brain MRE images.

### B. Validation of Modulus Estimation using TWENN

#### 1) Test Data Set

The test data set contained $10^7$ examples for model evaluation. Wavefields with different signal-to-noise ratios (SNRs) and complex wavenumbers were generated using the TWE model, with a patch size of 3×3 and a resolution of 3 mm × 3 mm at 60Hz. SNRs were distributed uniformly from 12 dB to 38 dB, real normalized wavenumbers were uniformly distributed from 0.35s/m to 1.35s/m, while imaginary normalized wavenumbers were uniformly distributed between 0s/m and 0.28s/m. Estimation of $\widetilde{k'}$ and $\widetilde{k''}$ at different SNRs and complex

| Application | Data Set | Details of data | Data Source | Method Proposed | Methods for Comparison | Evaluation Metrics |
|---|---|---|---|---|---|---|
| Phase unwrapping | Simulated Liver | 2mm×2mm; 42Hz; RO | BIOQIC | Dual-DC | SG | Mean error |
| | Brain MRE | 3mm×3mm×3mm; 50Hz; RO | 3T scanner | | LBE | Interlayer continuity |
| Modulus estimation | Test data | 3mm×3mm; 60Hz; SS | TWE model | TWENN (2D) | DI (2D) | Mean error |
| | Simulated phantom | 1.5mm×1.5mm; 60,80,100Hz; SS | COMSOL | | k-MDEV | RMSE |
| | Simulated brain | 1.5mm×1.5mm×1.5mm; 24, 28, 32, 36, 40, 44, 48, 52, 56, 60Hz; RO, PE, SS | BIOQIC | TWENN (3D) | MDEV (BIOQIC) | |
| | Simulated liver | 2mm×2mm×2mm 30, 36, 42, 48Hz; RO, PE, SS | | | | |
| Pipeline performance | Phantom | 1.5mm×1.5mm; 30, 40, 50, 60, 70, 80, 90, 100Hz; SS | BIOQIC | Dual-DC + TWENN (2D) | LBE + k-MDEV | Regional mean |
| | Normal brain | 2mm×2mm; 20, 25, 30, 35, 40, 45Hz; RO, PE, SS | | | | Match with structural image |
| | Normal liver | 2.7mm×2.7mm; 30, 40, 50, 60Hz; RO, PE, SS | 1.5T scanner | | PG + MDEV (BIOQIC) | Regional mean |
| | Hepatic siderosis & Cirrhosis | | | | | |
| | Liver tumor | | | | | |
| | Brain tumor | 3mm×3mm×3mm; 30, 40, 50Hz; RO, PE, SS | 3T scanner | Dual-DC + TWENN (3D) | | CNR |

wavenumbers were compared with those from conventional DI (fitting window=3×3).

### 2) Simulated Dataset

The size of the simulated phantom using COMSOL (COMSOL, Stockholm, Sweden) had two circular inclusions of 10 mm radius with complex shear moduli of 4 + i0.48 and 6 + i0.84 kPa. The complex shear modulus of background was 2 + i0.2 kPa. These values served as GT. A 28dB SNR of Gaussian noise was added to the wave images.

Publicly available MRE data for simulations of the liver and brain (https://bioqic-apps.charite.de/) were used for evaluation. Both were 3D, FE simulation models based on COMSOL. Model geometries were constructed from segmented images of a healthy human. Linear viscoelastic models were used and the tissue properties were assigned based on previously reported values [36], [37]. The data set contained a 3D multi-frequency and multi-direction wavefield with known $G'$ (GT). The liver dataset had a resolution of 2 mm × 2 mm × 4 mm with 4 frequency components (30, 36, 42, 48 Hz). The brain data set had a resolution of 1.5 mm × 1.5 mm × 1.5 mm, and 10 frequency components (24, 28, 32, 36, 40, 44, 48, 52, 56, 60 Hz).

Root mean squared error (RMSE) of estimated modulus $G_{est}'$ and the structures were compared with k-MDEV and MDEV [38]. RMSE is defined as:

$$\text{RMSE} = \sqrt{\frac{\|G_{est}' - G_{GT}'\|_2^2}{N}} \tag{18}$$

### C. Validation for the Proposed Pipeline

#### 1) Comparison with Other Pipelines

Both phantom and human MRE experiment images were used to evaluate the performance of the proposed pipeline for MRE image processing.

The proposed pipeline contains a cascading of phase unwrapping method using Dual-DC and TWENN for modulus estimation. Comparisons were made with two state-of-the-art multi-frequency MRE processing pipelines available on a public platform [38]: LBE and k-MDEV, PG and MDEV.

#### 2) Phantom MRE

For phantom validation, the multi-frequency phantom dataset containing four cylindrical inclusions was used [20]. The moduli estimated by each method were averaged in the z-direction. Regional mean values and ground truth of $G'$ and $G''$ were compared.

### 3) Brain MRE

Brain MRE images of a patient with a meningioma were acquired using a 3T scanner (uMR790, United Imaging Healthcare, Shanghai, China) using an electromagnetic actuator [34]. MRE was implemented using a single-shot spin-echo echo-planar imaging sequence (TR/TE=4000/65ms) with a motion-encoding gradient of 40 mT/m and 8 phase offsets. Contrast-to-noise ratio (CNR) of $G'$ with respect to the tumor region was used for evaluation:

$$\text{CNR} = \frac{2\left(\overline{G_{bkg}'} - \overline{G_{tumor}'}\right)^2}{\sigma_{bkg}^2 + \sigma_{tumor}^2}, \tag{19}$$

where $\overline{G_{bkg}'}$ and $\overline{G_{tumor}'}$ are the mean value of shear modulus, $\sigma_{bkg}$ and $\sigma_{tumor}$ are the standard deviation of shear modulus, for background tissue and tumor tissue, respectively. The tumor region was delineated from the T1 structural image, and the background neighborhood was delineated as the region of one tumor radius outside of the tumor.

A public data set of normal brain MRE images (https://bioqic-apps.charite.de/) was also used for evaluation by comparing the estimated modulus map and the structural image.

### 4) Liver MRE

Liver MRE images from a healthy volunteer, a patient with a liver tumor, and a patient with hepatic siderosis & cirrhosis were acquired using a 1.5T pneumatic MRE system [39] (Magnetom Aera, Siemens, Erlangen, Germany, BIOQIC). CNR of $G'$ was used to evaluate the modulus estimation of liver tumor data. For normal liver data and iron-deposited cirrhotic data, the mean values of $G'$ of circled liver tissue were compared.

### D. Implementation Details for Dual-DC and TWENN

For all datasets unwrapped in this paper, the same hyperparameters of Dual-DC were used (learning rate = 0.005, $\lambda = 1000$, the number of iterations = 4000). The networks were trained using the ADAM algorithm with a batch size of 500 and a step number of 12000. Before the complex wave field was introduced into the network, complex normalization was performed based on the maximum absolute value of all data points. In this way, the absolute value of the input data was normalized between 0 to 1. Due to the fact that physiological noise such as vascular pulsation may pollute *in vivo* measurement of brain and liver [40], [41], a relatively larger noise intensity $b$ of 0.3 was added to the training sets, compared with that of 0.001 for the remaining non-living measurement dataset (all simulation data and phantom data). If the wavefield was continuous between imaging slices, a 3×3×3 patch was used. Otherwise, 3×3 patch was used.

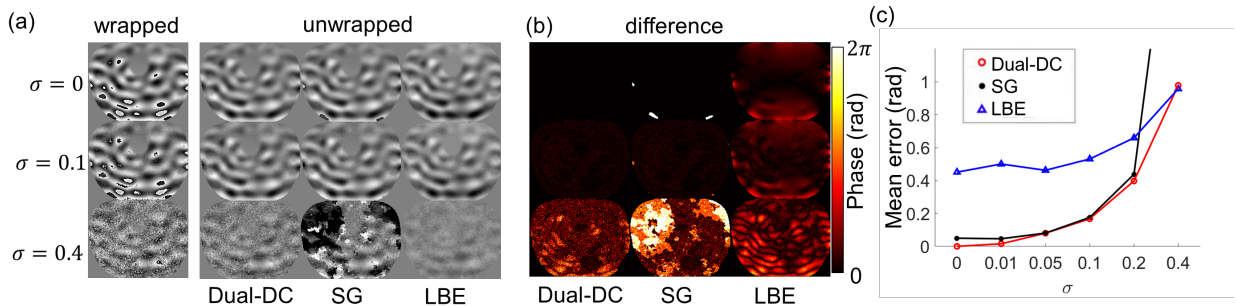

Fig. 2. (a) A comparison of the performance of the three unwrapping methods at varying noise levels. (b) The differences between the ground truth and the outputs of three methods. (c) Mean error at different noise levels.

The proposed methods were implemented with PyTorch 1.12.0 and CUDA 11.6 on an Ubuntu 20.04 LTS (64-bit) operating system equipped with an AMD Ryzen 9 5950x central processing unit (CPU) and NVIDIA RTX 3080Ti graphics processing unit (GPU, 12 GB memory).

# IV. RESULTS

## A. Phase unwrapping test

### 1) Simulation test

Results of unwrapping simulation data showed the proposed optimization-based algorithm using dual-DC could unwrap phases with a complex Gaussian noise level up to $\sigma = 0.4$ without observing the offset of the background phase (Fig. 2a). However, SG did not perform unwrap at some complex edges without noise, and a large number of unwrapping failures were observed with increased noise levels (Fig. 2a, b). Although LBE had no obvious unwrapping failures, it introduced background offset even without noise (Fig. 2b). In contrast, an optimization-based algorithm using Dual-DC had better performance at different noise levels (Fig. 2c).

### 2) In vivo dataset test

It was observed that SG was prone to unwrapping failure. The background offset introduced by the LBE could be seen in both coronal and sagittal views leading to an obvious 3D wavefield discontinuity. All these methods performed unwrapping at the transverse layers. However, an optimization-based algorithm using Dual-DC still ensured continuity at coronal and sagittal views (Fig. 3). The absolute differences also showed that Dual-DC produced the smoothest phase.

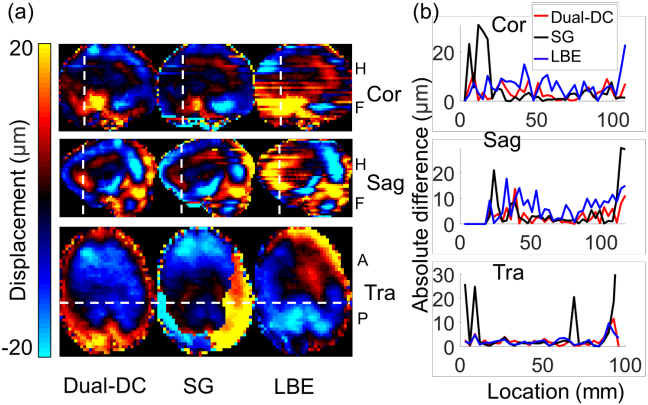

Fig. 3. (a) comparison of displacement field extracted from brain MRE images. (b) The corresponding absolute differences of displacement at cross sections marked by the white dotted line.

## B. Modulus estimation test

### 1) Simulation data set

A comparison of inversion between DI and TWENN with different complex wavenumbers and signal-to-noise ratios (SNR) in 2D wavefield (patch size=3×3) showed TWENN performed better than that of DI in estimating $\widetilde{k'}$ and $\widetilde{k''}$. As SNR decreased, the advantages of TWENN over DI increased (Fig. 4a, b). Furthermore, the mean errors increased with the value of modulus, indicating an upper limit of estimation using TWENN. The upper limit, however, is higher than that of DI as shown in the figure.

Results of the estimated shear moduli showed TWENN could recover the inclusions better than other methods (Fig. 5). Values of RMSE showed TWENN was the best at estimating complex shear moduli of the inclusions in most cases (TABLE II).

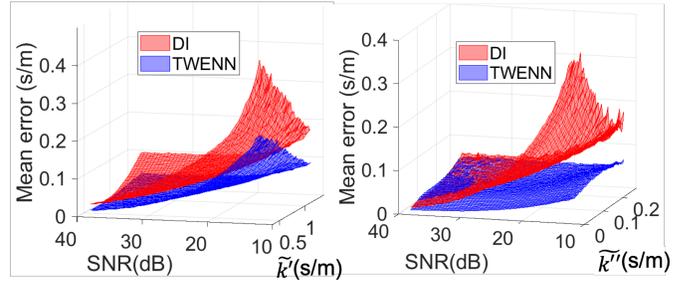

Fig. 4. Distribution of mean errors of estimating $\widetilde{k'}$ (left) and $\widetilde{k''}$ (right) for different SNR values and normalized complex wavenumber.

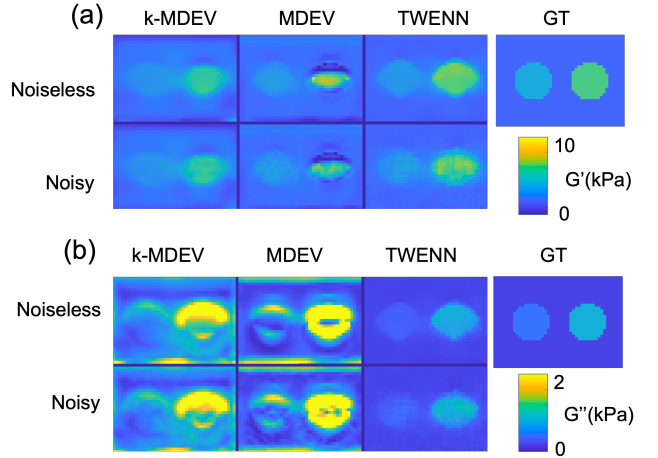

Fig. 5. Comparisons of estimated (a) $G'$ and (b) $G''$ maps using different methods based on phantom simulation. 28dB of noise was added to the simulated wave images.

### 2) Simulation test of brain and liver

Based on the simulated complex wavefield of the brain and liver, it was observed that TWENN had the best estimation

## TABLE II

Comparisons of RMSE in estimating the complex shear moduli of a simulated phantom, simulated brain and liver. Values of and CNR are also compared for imaging brain and liver tumors. The best results are shown in bold fonts.

| Method | RMSE$_{G'}$+RMSE$_{G''}$i (kPa) | | | RMSE (kPa) | | CNR | |
|---|---|---|---|---|---|---|---|
| | BK | Inc#1 | Inc#2 | Brain simulation | Liver simulation | Brain tumor | Liver tumor |
| k-MDEV | 0.65+0.67i | 0.91+0.32i | 1.59+1.38i | 3.65 | 2.4 | 826 | 357 |
| MDEV | **0.43**+0.59i | 0.89+0.54i | 3.52+2.28i | 3.94 | 2.67 | 16 | 79 |
| TWENN | 0.60+**0.10i** | **0.83+0.13i** | **1.12+0.19i** | **3.34** | **2.06** | **3069** | **3906** |

accuracy with RMSEs of 3.34 and 2.06kPa (TABLE II). The other two algorithms did not perform as well as TWENN probably due to the strong reflection of waves or the use of derivatives, which might produce artifacts for boundaries (Fig. 6).

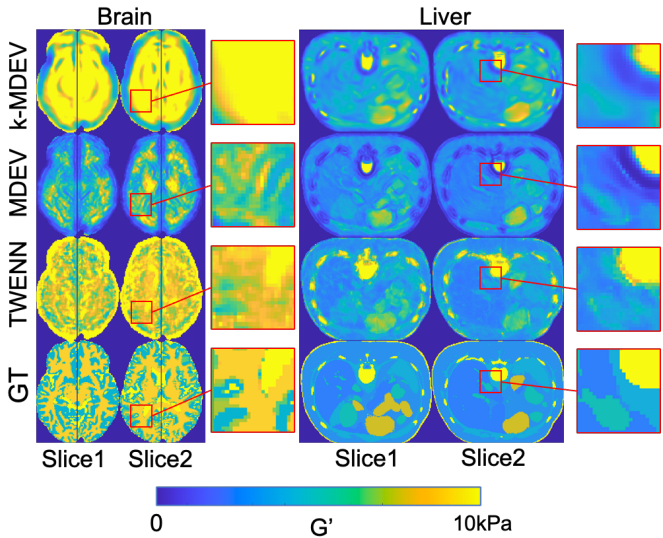

Fig. 6. A comparison of estimated $G'$ using different algorithms for simulated brain and liver. Regions were magnified to illustrate the anatomical features of $G'$.

## C. Performance Evaluation

### 1) Phantom MRE

Using multi-frequency MRE images of phantom, estimated $G'$ and $G''$ values showed the proposed pipeline had the best performance in recovering the inclusion structures (Fig. 7a). Algorithms using the Laplace operator could not recover the structures well, probably because phantom images contained various noise and wave deflections. With longer wavelengths and lower SNR in regions with relatively high modulus, these algorithms were prone to underestimate the high modulus. The proposed pipeline produced the closest modulus estimation to the ground truth for both $G'$ and $G''$ at most cases (Fig. 7b).

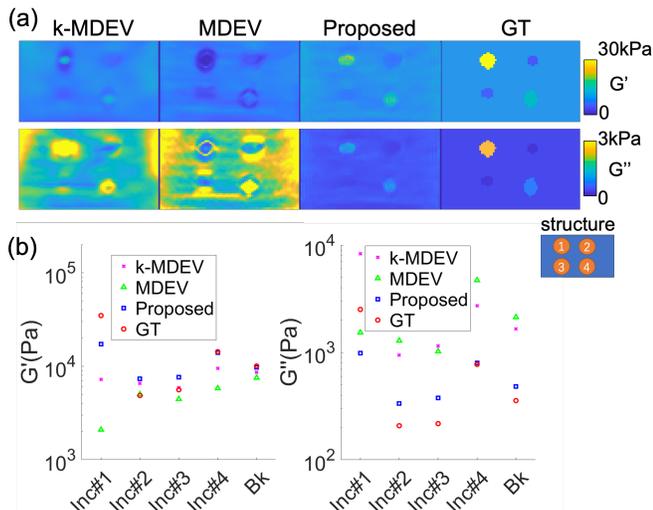

Fig. 7. (a) A comparison of estimated $G'$ and $G''$ maps of gelatin phantom using different processing methods. (b) A comparison of mean values of $G'$ and $G''$ of each method. The longitudinal axis is displayed in a logarithmic form.

### 2) Tumor MRE

For both brain and liver tumors, the proposed pipeline had better reconstruction of boundaries (Fig. 8a, b), with more details matching that of T1 images and the highest values of CNR (TABLE II). This showed the ability of the proposed pipeline to reconstruct the structural details.

### 3) Normal brain MRE

For the normal human brain dataset, $G'$ values estimated from MDEV were relatively low (<1kPa). This was probably affected by noise. Both the k-MDEV and the proposed pipeline could estimate shear modulus maps which match the structure image. k-MDEV introduced significant boundary artifacts which is consistent with the results in Fig. 6. As indicated by the white arrow in Slice 1, the brain's midline was better reconstructed by the proposed pipeline than that of other algorithms (Fig. 8c). The white arrow of Slice 2 (Fig. 8c) showed k-MDEV estimated relatively low modulus (<0.3 kPa), but the proposed method provided a proper estimate (~1kPa). This dataset demonstrated the capability of the proposed pipeline to reconstruct anatomical features.

### 4) Liver dataset test

For liver MRE (Fig. 8e, f), the estimated $G'$ values of the normal liver from k-MDEV, MDEV, and the proposed pipeline were 1.62, 1.31, and 1.97 kPa, respectively. These values were all in line with the reported normal range [42]. However, in the case of cirrhotic with iron deposition, the estimated $G'$ values were 1.70 and 1.09 for k-MDEV and MDEV, respectively. This was likely caused by the iron deposition that resulted in a low SNR. The value of 3.13 kPa estimated by the proposed pipeline fell within the reported range [42].

## V. Discussion

A pipeline for estimating complex moduli from MRE images is proposed in this study. The processing pipeline contains two major steps: displacement extraction and modulus estimation. For displacement extraction, an optimization-based method with Dual-DC terms is used for both calculating principal components and phase unwrapping. For modulus estimation, a complex neural network using covariance as input is proposed based on the TWE model.

Since discontinuities of unwrapped phase images can introduce serious artifacts of the derived modulus distribution, accurate phase unwrapping is an important prerequisite. Complex or noisy wavefields are difficult to deal with using search-like or discrete greedy update phase unwrapping techniques, such as the SG algorithm. Laplacian-based estimation (LBE) can perform unwrapping well in noisy cases, but extra background offset noise could be introduced. The Dual-DC based continuous optimization can deal with unwrapping of complex wrapped and noisy phase without introducing extra phase offset. The proposed method for unwrapping uses phase gradient information, which is more adaptable to complex scenarios than discrete unwrapping. Compared to LBE, Dual-DC does not introduce new background noise. Phase images were unwrapped successfully for wavefield data from phantom, brain and liver, using the same set of hyperparameters. This verifies the generalizability of the proposed algorithm. In addition, the computation time for MRE images of the whole brain (3 frequencies and 3 directions) was about 3 minutes.

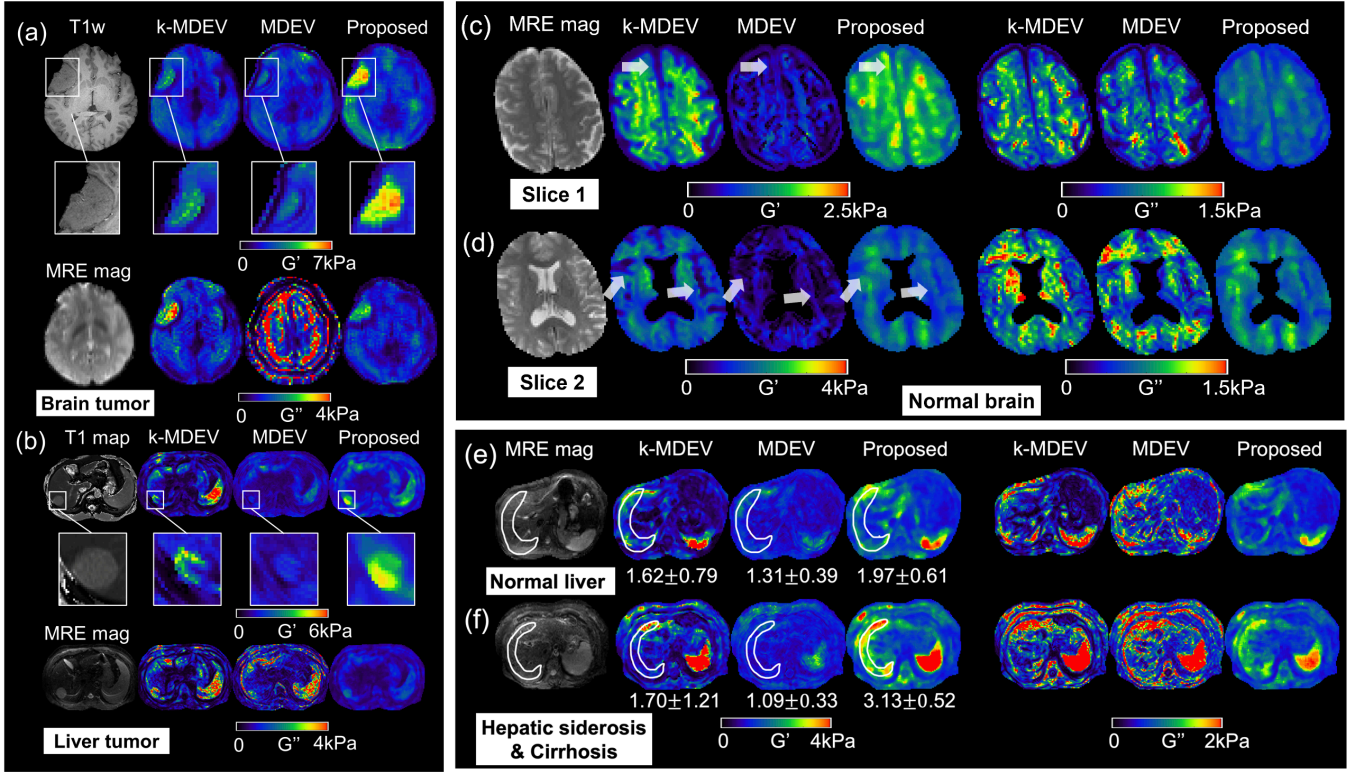

**Fig. 8.** T1-weighted (T1w) images, magnitude images of MRE, $G'$ and $G''$ maps estimated from different algorithms were compared for (a) brain tumor, (b) liver tumor, and (c, d) two difference slices of normal brain images. Magnified views of the tumor region were also provided. Magnitude images of MRE, $G'$ and $G''$ maps estimated from (e) a healthy volunteer and (f) a patient with hepatic siderosis & cirrhosis were compared. Regions of interest were delineated with white line, and the corresponding values of mean±standard deviation of $G'$ were provided.

For modulus inversion, where noise robustness, structural resilience and computational complexity impose contradicting constraints, TWENN manages to reach a tradeoff among these three metrics. In terms of noise resilience, TWENN does not use explicit differentiation computation, which also provides structural resolution benefits by reducing potential edge artifacts. Except for the LFE-type and NLI-type methods [9], [25], [23], [26], most of current algorithms rely on differential computation, which can be sensitive to noise. By adding noise to the training set, TWENN can simultaneously achieve both denoising and inversion, improving its performance at regions with low SNR. For example, TWENN can estimate the modulus of cirrhotic tissue at low SNR regions induced by iron deposition (Fig. 8f). TWENN has the potential to improve the clinical performance of liver MRE due to the fact that iron deposition is a significant cause of liver MRE failure [42].

Considering structural resilience, a small patch of 3×3 or 3×3×3 pixels is applied so that TWENN can still distinguish details of the image objects even if local homogeneity is assumed. Benefiting from the absence of differential calculations, no significant artifacts are observed at the structural edges. Compare with MDEV and k-MDEV, both of which apply differential operations, TWENN has better performance in preserving anatomical details on both simulated and in vivo liver and brain datasets (Fig. 6, Fig. 8).

If the estimation kernel is chosen in a trial-and-error fashion, it is difficult to ensure the noise robustness of small patches.

Therefore, neural networks are used here to establish the nonlinear mapping between noisy wavefields and complex wavenumbers. Even training without noise, TWENN still has better noise robustness than conventional DI using a small kernel (Fig. 4). The dimension of the kernel also affects the inversion. The use of a small sized patch will set an upper limit to the estimated modulus. Using dejittering method, it has been shown that interslice phase inconsistencies introduced in signal acquisition can be removed, providing robust 2D inversions, especially in the abdomen . Here, with improved phase unwrapping, the smooth and physically accurate phase can result in better modulus estimation from 3D inversion.

TWENN uses neural networks to pre-learn wavefield structure information for inversion. The time of training data generation and network training using a desktop PC did not exceed 3 minutes. The trained network can be directly used to process MRE images with the same vibration frequency and resolution. In this study, the 3D multi-directional and multi-frequency inversion time of a whole brain did not exceed 15 seconds, showing its potential to be deployed clinically. This is because that all these methods use multi-frequency fusion, which is not used by other leading FE-based approaches like NLI. Furthermore, inverse FE-based algorithms require iterative FE computation updates [25], [37], [43] which may result in computations ten times or higher than TWENN, MEDV, and k-MEDV.

The modulus inversion of MRE is equivalent to estimating the wavenumber for an array of wave points, which can be considered as a variant of the Direction of Arrival (DOA) problem in array signal processing. Strategies commonly used in DOA solution problems [30] can be drawn on, where second-order moments can effectively reveal harmonic information. The possibility of estimating modulus from an array signal processing perspective was validated in our previous work [44].

From the perspective of generalizability, the training data set in TWENN is produced using a wave equation. Theoretically, any local wavefield can be represented using the TWE model, including reflected standing waves. Therefore, the training set can cover most of the wave propagation conditions, ensuring the generalizability of TWENN. Existing neural network training modulus estimation methods [28], [29] are usually trained using FE simulation for a limited number of scenarios, which could hamper their generalization performance. In addition, most of the elastography processing pipelines used various complex filters for either wave extraction or denoising [20], [21], [22]. However, the proposed pipeline does not use any filters other than a median filter with fixed parameters. Therefore, complicated parameter tuning procedures are prevented, improving its generalization performance.

Although multi-frequency and multi-directional information is used to improve inversion performance [2], TWENN also supports single-frequency and single-direction inversion. In addition to $G'$ and $G''$ values, shear wave speed and penetrating rate that are closely related with stiffness and damping [2] can also be estimated.

Limitations of this study include assumptions of local homogeneity, isotropy, and linear viscoelasticity, which could be addressed by more realistic FE simulation based NLI [9], [23], [24], [25], [45], [46], and neural network based inversion [29], [47], [48]. Although TWENN can estimate a relatively smoother viscosity distribution from the noisy wavefield, only simple phantom cases were used to validate it. In this study, limited human imaging data were used for validation. Future work includes applying the proposed pipeline to a larger cohort of clinical data sets such as neurological and liver diseases.

The unwrapping method proposed in this paper is readily transferable to other fields where unwrapping is required. The inversion method can also be applied to other elastic imaging modalities, such as optical elastography and ultrasound elastography where wave equations apply. The TWE model can be modified in terms of specific anisotropic and dispersive material models, in order to solve more complex modulus inversion problems. Future work includes estimating properties of transversely isotropic and frequency-dependent materials within the TWENN framework. The potential of using this data-driven method for obtaining inverse operators to solve other inverse problems such as electrical resistance tomography needs to be further explored.

## APPENDIX

### TRAINING DATA GENERATION

The training data set was generated using the following parameter settings:

1) The number $M$ of traveling waves 1,2,3,4,5,6,7,8 was set to be distributed at ratios of 6:4:3:2:1:1:1:1.
2) The complex amplitude $|a_m|$ was uniformly distributed in $[0,1]$ and $\text{angle}(a_m)$ was uniformly distributed in $[0,2\pi]$.
3) The unit vector of the wave propagation direction is $\widehat{n_m} = [x, y, z]$.
   a) In 2D cases, $x = \cos(\theta), y = \sin(\theta), z = 0, \theta$ was uniformly distributed in $[0,2\pi]$.
   b) In 3D cases, $z$ is uniformly distributed in $[-1,1]$, $\theta$ was uniformly distributed in $[0,2\pi]$, $x = \sqrt{1-z^2}\cos(\theta)$, $y = \sqrt{1-z^2}\sin(\theta)$.
4) $k'$ and $k''$ were uniformly distributed within the preset range.

## ACKNOWLEDGMENT

We thank Prof. Ingolf Sack from Charite, Germany for providing part of the testing data sets and helpful discussions.

## REFERENCES

[1] S. K. Venkatesh and R. L. Ehman, Eds., *Magnetic Resonance Elastography*. New York, NY: Springer New York, 2014. doi: 10.1007/978-1-4939-1575-0.

[2] I. Sack, "Magnetic resonance elastography from fundamental soft-tissue mechanics to diagnostic imaging," *Nat. Rev. Phys.*, Nov. 2022, doi: 10.1038/s42254-022-00543-2.

[3] H. Morisaka *et al.*, "Magnetic resonance elastography is as accurate as liver biopsy for liver fibrosis staging: MRE Is Equivalent to Liver Biopsy," *J. Magn. Reson. Imaging*, vol. 47, no. 5, pp. 1268–1275, May 2018, doi: 10.1002/jmri.25868.

[4] Y. Shi *et al.*, "Use of magnetic resonance elastography to gauge meningioma intratumoral consistency and histotype," *NeuroImage Clin.*, vol. 36, p. 103173, 2022, doi: 10.1016/j.nicl.2022.103173.

[5] P. Garteiser *et al.*, "MR elastography of liver tumours: value of viscoelastic properties for tumour characterisation," *Eur. Radiol.*, 2012, doi: 10.1007/s00330-012-2474-6.

[6] A. Lipp, C. Skowronek, A. Fehlner, K.-J. Streitberger, J. Braun, and I. Sack, "Progressive supranuclear palsy and idiopathic Parkinson's disease are associated with local reduction of in vivo brain viscoelasticity," *Eur. Radiol.*, vol. 28, no. 8, pp. 3347–3354, Aug. 2018, doi: 10.1007/s00330-017-5269-y.

[7] S. Hirsch, J. Braun, and I. Sack, *Magnetic Resonance Elastography*. 2017.

[8] E. Barnhill, P. Kennedy, C. L. Johnson, M. Mada, and N. Roberts, "Real-time 4D phase unwrapping applied to magnetic resonance elastography: Real-Time 4D Phase Unwrapping," *Magn. Reson. Med.*, vol. 73, no. 6, pp. 2321–2331, Jun. 2015, doi: 10.1002/mrm.25332.

[9] D. Fovargue, D. Nordsletten, and R. Sinkus, "Stiffness reconstruction methods for MR elastography: Stiffness reconstruction methods for MR elastography," *NMR Biomed.*, vol. 31, no. 10, p. e3935, Oct. 2018, doi: 10.1002/nbm.3935.

[10] Y. Feng, E. H. Clayton, R. J. Okamoto, J. Engelbach, P. V. Bayly, and J. R. Garbow, "A longitudinal magnetic resonance elastography study of murine brain tumors following radiation therapy," *Phys. Med. Biol.*, vol. 61, no. 16, pp. 6121–6131, 2016, doi: 10.1088/0031-9155/61/16/6121.

[11] G. Z. Yang, P. Burger, P. J. Kilner, S. P. Karwatowski, and D. N. Firmin, "Dynamic range extension of cine velocity measurements using motion registered spatiotemporal phase unwrapping," *J. Magn. Reson. Imaging*, vol. 6, no. 3, pp. 495–502, May 1996, doi: 10.1002/jmri.1880060313.

[12] M. A. Herráez, D. R. Burton, M. J. Lalor, and M. A. Gdeisat, "Fast two-dimensional phase-unwrapping algorithm based on sorting by reliability following a noncontinuous path," *Appl. Opt.*, vol. 41, no. 35, p. 7437, Dec. 2002, doi: 10.1364/AO.41.007437.

[13] V. V. Volkov and Y. Zhu, "Deterministic phase unwrapping in the presence of noise," *Opt. Lett.*, vol. 28, no. 22, p. 2156, Nov. 2003, doi: 10.1364/OL.28.002156.

[14] K.-J. Streitberger *et al.*, "High-Resolution Mechanical Imaging of Glioblastoma by Multifrequency Magnetic Resonance Elastography," *PLoS ONE*, vol. 9, no. 10, p. e110588, Oct. 2014, doi: 10.1371/journal.pone.0110588.

[15] T. E. Oliphant, A. Manduca, R. L. Ehman, and J. F. Greenleaf, "Complex-valued stiffness reconstruction for magnetic resonance elastography by algebraic inversion of the differential equation," *Magn. Reson. Med.*, vol. 45, no. 2, pp. 299–310, Feb. 2001, doi: 10.1002/1522-2594(200102)45:2<299::AID-MRM1039>3.0.CO;2-O.

[16] H. Knutsson, C.-F. Westin, and G. Granlund, "Local multiscale frequency and bandwidth estimation," in *Proceedings of 1st International Conference on Image Processing*, Austin, TX, USA: IEEE Comput. Soc. Press, 1994, pp. 36–40. doi: 10.1109/ICIP.1994.413270.

[17] S. Mohammed *et al.*, "Model-based quantitative elasticity reconstruction using ADMM," *IEEE Trans. Med. Imaging*, pp. 1–1, 2022, doi: 10.1109/TMI.2022.3178072.

[18] H. Dong, R. Ahmad, R. Miller, and A. Kolipaka, "MR elastography inversion by compressive recovery," *Phys. Med. Biol.*, vol. 66, no. 16, p. 165001, Aug. 2021, doi: 10.1088/1361-6560/ac145a.

[19] L. Hu and X. Shan, "Enhanced complex local frequency elastography method for tumor viscoelastic shear modulus reconstruction," *Comput. Methods Programs Biomed.*, vol. 195, p. 105605, Oct. 2020, doi: 10.1016/j.cmpb.2020.105605.

[20] S. Papazoglou, S. Hirsch, J. Braun, and I. Sack, "Multifrequency inversion in magnetic resonance elastography," *Phys. Med. Biol.*, vol. 57, no. 8, pp. 2329–2346, Apr. 2012, doi: 10.1088/0031-9155/57/8/2329.

[21] H. Tzschätzsch *et al.*, "Tomoelastography by multifrequency wave number recovery from time-harmonic propagating shear waves," *Med. Image Anal.*, vol. 30, pp. 1–10, May 2016, doi: 10.1016/j.media.2016.01.001.

[22] E. Barnhill *et al.*, "Nonlinear multiscale regularisation in MR elastography: Towards fine feature mapping," *Med. Image Anal.*, vol. 35, pp. 133–145, Jan. 2017, doi: 10.1016/j.media.2016.05.012.

[23] E. Van Houten, K. Paulsen, M. Miga, F. Kennedy, and J. Weaver, "An overlapping subzone technique for MR-based elastic property reconstruction," *Magn. Reson. Med.*, vol. 42, no. 4, pp. 779–786, Oct. 1999, doi: 10.1002/(SICI)1522-2594(199910)42:4<779::AID-MRM21>3.3.CO;2-Q.

[24] B. Babaei *et al.*, "Magnetic Resonance Elastography Reconstruction for Anisotropic Tissues," *Med. IMAGE Anal.*, vol. 74, Dec. 2021, doi: 10.1016/j.media.2021.102212.

[25] M. Honarvar, R. S. Sahebjavaher, R. Rohling, and S. E. Salcudean, "A Comparison of Finite Element-Based Inversion Algorithms, Local Frequency Estimation, and Direct Inversion Approach Used in MRE," *IEEE Trans. Med. IMAGING*, vol. 36, no. 8, pp. 1686–1698, Aug. 2017, doi: 10.1109/TMI.2017.2686388.

[26] M. D. J. McGarry *et al.*, "Multiresolution MR elastography using nonlinear inversion," *Med. Phys.*, vol. 39, no. 10, pp. 6388–6396, Oct. 2012, doi: 10.1118/1.4754649.

[27] L. V. Hiscox *et al.*, "Standard-space atlas of the viscoelastic properties of the human brain," *Hum. BRAIN Mapp.*, vol. 41, no. 18, pp. 5282–5300, Dec. 2020, doi: 10.1002/hbm.25192.

[28] M. C. Murphy, A. Manduca, J. D. Trzasko, K. J. Glaser, J. Huston, and R. L. Ehman, "Artificial neural networks for stiffness estimation in magnetic resonance elastography: Neural Network Inversion for MRE," *Magn. Reson. Med.*, vol. 80, no. 1, pp. 351–360, Jul. 2018, doi: 10.1002/mrm.27019.

[29] J. M. Scott *et al.*, "Artificial neural networks for magnetic resonance elastography stiffness estimation in inhomogeneous materials," *Med. Image Anal.*, vol. 63, p. 101710, Jul. 2020, doi: 10.1016/j.media.2020.101710.

[30] A. Baghani, S. Salcudean, M. Honarvar, R. S. Sahebjavaher, R. Rohling, and R. Sinkus, "Travelling Wave Expansion: A Model Fitting Approach to the Inverse Problem of Elasticity Reconstruction," *IEEE Trans. Med. Imaging*, vol. 30, no. 8, pp. 1555–1565, Aug. 2011, doi: 10.1109/TMI.2011.2131674.

[31] C. Trabelsi *et al.*, "Deep Complex Networks." arXiv, Feb. 25, 2018. Accessed: Oct. 25, 2022. [Online]. Available: http://arxiv.org/abs/1705.09792

[32] M. Arjovsky, A. Shah, and Y. Bengio, "Unitary Evolution Recurrent Neural Networks," *33rd Int. Conf. Mach. Learn.*, vol. 48, p. 9, 2016.

[33] B. Ottersten, P. Stoica, and R. Roy, "Covariance Matching Estimation Techniques for Array Signal Processing Applications," *Digit. Signal Process.*, vol. 8, no. 3, pp. 185–210, Jul. 1998, doi: 10.1006/dspr.1998.0316.

[34] S. Qiu *et al.*, "An electromagnetic actuator for brain magnetic resonance elastography with high frequency accuracy," *NMR Biomed.*, vol. 34, no. 12, Dec. 2021, doi: 10.1002/nbm.4592.

[35] F. Dittmann, S. Hirsch, H. Tzschätzsch, J. Guo, J. Braun, and I. Sack, "In vivo wideband multifrequency MR elastography of the human brain and liver: In Vivo Multifrequency wMRE of the Human Brain and Liver," *Magn. Reson. Med.*, vol. 76, no. 4, pp. 1116–1126, Oct. 2016, doi: 10.1002/mrm.26006.

[36] C. Ariyurek, B. Tasdelen, Y. Z. Ider, and E. Atalar, "SNR Weighting for Shear Wave Speed Reconstruction in Tomoelastography," *NMR Biomed.*, vol. 34, no. 1, Jan. 2021, doi: 10.1002/nbm.4413.

[37] E. Barnhill, M. Nikolova, C. Ariyurek, F. Dittmann, J. Braun, and I. Sack, "Fast Robust Dejitter and Interslice Discontinuity Removal in MRI Phase Acquisitions: Application to Magnetic Resonance Elastography," *IEEE Trans. Med. Imaging*, vol. 38, no. 7, pp. 1578–1587, Jul. 2019, doi: 10.1109/TMI.2019.2893369.

[38] T. Meyer *et al.*, "Comparison of inversion methods in MR elastography: An open-access pipeline for processing multifrequency shear-wave data and demonstration in a phantom, human kidneys, and brain," *Magn. Reson. Med.*, vol. 88, no. 4, pp. 1840–1850, Oct. 2022, doi: 10.1002/mrm.29320.

[39] F. Dittmann *et al.*, "Tomoelastography of the abdomen: Tissue mechanical properties of the liver, spleen, kidney, and pancreas from single MR elastography scans at different hydration states: Tomoelastography of the Abdomen," *Magn. Reson. Med.*, vol. 78, no. 3, pp. 976–983, Sep. 2017, doi: 10.1002/mrm.26484.

[40] A. J. Hannum, G. McIlvain, D. Sowinski, M. D. J. McGarry, and C. L. Johnson, "Correlated noise in brain magnetic resonance elastography," *Magn. Reson. Med.*, vol. 87, no. 3, pp. 1313–1328, Mar. 2022, doi: 10.1002/mrm.29050.

[41] M. Shahryari *et al.*, "Reduction of breathing artifacts in multifrequency magnetic resonance elastography of the abdomen," *Magn. Reson. Med.*, vol. 85, no. 4, pp. 1962–1973, Apr. 2021, doi: 10.1002/mrm.28558.

[42] M. Yin, K. J. Glaser, J. A. Talwalkar, J. Chen, A. Manduca, and R. L. Ehman, "Hepatic MR Elastography: Clinical Performance in a Series of 1377 Consecutive Examinations," *Radiology*, vol. 278, no. 1, pp. 114–124, Jan. 2016, doi: 10.1148/radiol.2015142141.

[43] I. M. Perreard *et al.*, "Effects of frequency-and direction-dependent elastic materials on linearly elastic MRE image reconstructions," *Phys. Med. Biol.*, vol. 55, no. 22, pp. 6801–6815, Nov. 2010, doi: 10.1088/0031-9155/55/22/013.

[44] S. Ma and Y. Feng, "A traveling-wave expansion based subspace search (TWESS) for shear modulus estimation using magnetic resonance elastography," presented at the 2021 International Society for Magnetic Resonance in Medicine (ISMRM2021), May 2022.

[45] M. McGarry, E. Van Houten, L. Solamen, S. Gordon-Wylie, J. Weaver, and K. Paulsen, "Uniqueness of poroelastic and viscoelastic nonlinear inversion MR elastography at low frequencies," *Phys. Med. Biol.*, vol. 64, no. 7, Apr. 2019, doi: 10.1088/1361-6560/ab0a7d.

[46] M. McGarry *et al.*, "Mapping heterogenous anisotropic tissue mechanical properties with transverse isotropic nonlinear inversion MR elastography," *Med. Image Anal.*, vol. 78, p. 102432, May 2022, doi: 10.1016/j.media.2022.102432.

[47] J. M. Scott *et al.*, "Impact of material homogeneity assumption on cortical stiffness estimates by MR elastography," *Magn. Reson. Med.*, vol. 88, no. 2, pp. 916–929, Aug. 2022, doi: 10.1002/mrm.29226.

[48] Z. Hou *et al.*, "Estimation of the mechanical properties of a transversely isotropic material from shear wave fields via artificial neural networks," *J. Mech. Behav. Biomed. Mater.*, vol. 126, p. 105046, Feb. 2022, doi: 10.1016/j.jmbbm.2021.105046.